\documentclass[aps,prl,twocolumn,superscriptaddress,nofootinbib]{revtex4-2}

\usepackage{amsmath,amssymb,graphicx,bm}
\usepackage{hyperref}
\usepackage{tikz}
\usepackage{pgfplots}
\pgfplotsset{compat=1.18}

\begin{document}

\title{Probing Dark Energy on the Moon}

\author{Alfredo Gurrola,$^1$ Robert J. Scherrer,$^1$ Oem Trivedi}
\affiliation{Department of Physics and Astronomy, Vanderbilt University, Nashville, TN, USA}

\begin{abstract}
The effective field theory (EFT) of cosmic acceleration provides a model-independent framework for describing dark energy and modified gravity, yet many of its defining operators remain weakly constrained by existing observations. We show that measurements of horizon-scale metric fluctuations with a lunar laser interferometer directly probe the kinetic sector of the EFT of dark energy, enabling constraints on operators governing scalar perturbation dynamics rather than background expansion. In particular, we demonstrate sensitivity to the EFT kinetic coefficient $M_2^4$ and the associated sound speed of dark energy, $c_s^2$. This establishes a qualitatively new observational handle on the microphysical consistency conditions of late-time acceleration models, allowing broad classes of EFT parameter space to be either discovered or excluded.
\end{abstract}

\maketitle

\section{Introduction}

Explaining the origin of the observed late–time cosmic acceleration remains one of the most profound open problems in fundamental physics. Although the standard $\Lambda$CDM paradigm provides an excellent phenomenological description of a broad range of cosmological observations, the physical nature of the component driving acceleration, dark energy, \cite{de1SupernovaSearchTeam:1998fmf} is still unknown. This ambiguity has motivated a vast landscape of theoretical proposals, including dynamical scalar fields, modified gravity theories, and scenarios inspired by quantum gravity \cite{de2Li:2012dt,de3Li:2011sd,de6Huterer:2017buf}.

A particularly powerful framework for organizing and comparing these possibilities is the effective field theory (EFT) of cosmic acceleration \cite{de5Frusciante:2019xia,eft1gubitosi2013effective,eft2linder2016effective,eft3bloomfield2013dark,eft4liang2023dark,eft5frusciante2014effective}. The EFT formalism provides a unifying and model–independent language in which all operators consistent with the symmetries of a quasi FLRW spacetime are systematically parameterized. In this approach, different dark energy and modified gravity models correspond to specific choices of time–dependent EFT coefficients. Observational constraints can therefore be mapped directly onto combinations of operators that encode the underlying microphysics, without committing to a particular ultraviolet completion.

Most current observational constraints focus on background quantities, such as the dark energy equation of state parameter $w$. However, widely different theories can yield nearly identical background expansion histories \cite{de7Vagnozzi:2021quy,de8Adil:2023ara,de10DiValentino:2020evt,de11Nojiri:2010wj,de12Nojiri:2006ri}. Within the EFT framework, this degeneracy arises because background evolution probes only a restricted subset of operators, leaving the kinetic and gradient structure of perturbations comparatively unconstrained. In particular, the sound speed of dark energy perturbations, $c_s^2$, which governs the propagation of scalar fluctuations and controls the clustering properties of dark energy, remains only weakly bounded.

The value of $c_s^2$ provides a direct discriminator between classes of models. Canonical scalar field scenarios generically predict $c_s^2 = 1$, implying that dark energy perturbations propagate luminally and remain smooth on sub-horizon scales. In contrast, more general constructions including coupled dark energy models \cite{cde1gumjudpai2005coupled,cde2gomez2020update,cde3barros2019kinetically,cde4xia2009constraint,cde5maccio2004coupled}, k-essence and other non-canonical theories \cite{nc1armendariz2001essentials,nc2malquarti2003new,nc3armendariz2005haloes,nc4ahn2009dark,nc5chimento2010dbi,nc6cai2016dark,nc7chattopadhyay2010interaction,nc8calcagni2006tachyon,nc9bagla2003cosmology,nc10copeland2005needed,nc11micheletti2010observational,nc12sheykhi2012tachyon,Kehayias:2019gir,Chiba:2009nh,Das:2006cm}, and broad subclasses of the EFT parameter space, naturally predict $c_s^2 \ll 1$, enabling significant clustering on cosmological scales. In some cases, superluminal propagation ($c_s^2 > 1$) can also arise at the level of the low–energy effective description. Consequently, measuring or constraining $c_s^2$ provides direct information about the kinetic operators in the EFT action and about the stability and consistency of the theory.

Despite its theoretical importance, the sound speed of dark energy has proven extremely challenging to constrain observationally. Existing bounds are derived primarily from integrated effects in the late–time Universe, such as the Integrated Sachs–Wolfe (ISW) effect \cite{csc1Hannestad:2005ak,csc2de2010measuring,csc3ballesteros2010dark,csc4linton2018variable,csc5bean2004probing,csc6eisenstein2005dark,csc7sergijenko2015sound,ss1yang2025clustering}. These probes depend on line–of–sight integrals over evolving gravitational potentials and are therefore sensitive only indirectly to the underlying perturbation dynamics. As a result, large regions of EFT parameter space, particularly those governing the kinetic structure and propagation speed of scalar perturbations, remain effectively unconstrained.

In a recent work \cite{soundspeed}, we proposed a qualitatively new observational route to the microphysics of cosmic acceleration by showing that a lunar interferometer setup like the Lunar Interferometer for Laser Astronomy (LILA) \cite{lilajani2025laser} can directly probe the dark energy sound speed $c_s^2$ through real time measurements of horizon-scale scalar gravitational potentials in an ultra-low frequency band inaccessible from Earth. Treating dark energy both as a general fluid and within the EFT of cosmic acceleration, we showed that the sound speed $c_s^2$ governs clustering on scales $k\lesssim aH/c_s$, inducing characteristic scale and time dependent modifications of the gravitational potentials $(\Phi,\Psi)$ that map directly into an interferometric strain signal. We developed a likelihood and Fisher-forecast framework based on mock strain spectra demonstrating that lunar interferometry can achieve discovery level sensitivity to clustering dark energy ($c_s^2\ll1$) or exclude broad classes of perturbative models.

In this work, we take that line of thought forward and propose a qualitatively new observational avenue wherein direct measurements of the real–time evolution of horizon–scale metric perturbations using LILA provide direct sensitivity to the operators in the EFT action that control the kinetic term and gradient energy of dark energy fluctuations, and by probing the time dependence of large scale gravitational potentials, a lunar–based interferometer can access precisely those sectors of parameter space that are invisible to traditional background and integrated probes. We show that this approach enables direct tests of the stability conditions, clustering behavior and propagation speed of dark energy perturbations, independent of assumptions about ultraviolet completions. In particular, constraints on the strain induced by evolving horizon–scale potentials can be translated into bounds on $c_s^2$ within the EFT framework. Our results therefore illustrate how next–generation precision measurements of metric fluctuations in the Solar System can open a new window onto the microphysics of cosmic acceleration.

\section{Effective Field Theory of Dark Energy}

A systematic and model–independent description of dark energy and modified gravity is provided by the effective field theory (EFT) of cosmic acceleration~\cite{de5Frusciante:2019xia}. In unitary gauge, where fluctuations of the dark energy scalar are absorbed into the metric, the action can be written as an expansion in operators consistent with time–dependent spatial diffeomorphism invariance. At lowest order in fluctuations, the action takes the form
\begin{multline}
S = \int d^4x \sqrt{-g} 
\Bigg[
\frac{M_*^2}{2}R - \Lambda(t) - c(t) g^{00} + \\\\ \frac{M_2^4(t)}{2} (\delta g^{00})^2 + \cdots
\Bigg],
\end{multline}
where $M_*^2$ is the (possibly time–dependent) Planck mass, $\Lambda(t)$ and $c(t)$ determine the background evolution, and $M_2^4(t)$ multiplies the leading quadratic operator in lapse perturbations.

Among these coefficients, $M_2^4(t)$ plays a particularly important physical role. While $\Lambda(t)$ and $c(t)$ are fixed by the requirement of reproducing the desired background expansion history, the operator $(\delta g^{00})^2$ does not contribute to the homogeneous dynamics at leading order. Instead, it controls the kinetic structure of scalar perturbations. In this sense, $M_2^4$ governs the inertia of dark energy fluctuations and directly determines their propagation and clustering properties.

In the minimal truncation where $(\delta g^{00})^2$ provides the dominant modification to the scalar sector, the sound speed of dark energy perturbations is
\begin{equation} \label{basecs_new}
c_s^2 = \frac{c(t)}{c(t) + 2M_2^4(t)}.
\end{equation}
This expression makes the physical interpretation transparent. When $M_2^4 = 0$, one recovers $c_s^2 = 1$, corresponding to a canonical scalar field with luminal propagation and negligible clustering on sub-horizon scales. For $M_2^4 > 0$, the sound speed is reduced, $c_s^2 < 1$, allowing dark energy perturbations to cluster more efficiently on large scales. In the regime $M_2^4 \gg c(t)$, the scalar sector becomes strongly kinetically screened: fluctuations acquire large inertia, their propagation speed becomes small, and the field can cluster significantly on horizon scales. 

Crucially, because $M_2^4$ enters only at quadratic order in perturbations, it leaves the background expansion history essentially unchanged. As a result, it is largely invisible to distance–based probes such as supernova luminosity distances or baryon acoustic oscillations. Existing constraints on $M_2^4$, and therefore on $c_s^2$, arise mainly from integrated effects in the evolution of gravitational potentials, most notably the late–time Integrated Sachs–Wolfe (ISW) effect. These constraints are indirect and remain comparatively weak, leaving broad regions of parameter space unconstrained.

The EFT framework can be extended beyond this minimal truncation. In full generality, additional operators consistent with the residual symmetries may be included, such as couplings between the lapse perturbation and the extrinsic curvature, or quadratic combinations of curvature fluctuations:
\begin{multline}
S = \int d^4x \sqrt{-g}\Bigg[
\frac{M_*^2(t)}{2}R - \Lambda(t) - c(t) g^{00}
+ \frac{M_2^4(t)}{2}(\delta g^{00})^2 \\
- \frac{m_3^3(t)}{2}\delta g^{00}\delta K
+ \mu_2^2(t)\left(\delta K^2 - \delta K_{ij}\delta K^{ij}\right)
+ \cdots
\Bigg],
\end{multline}
where $\delta K_{ij}$ denotes perturbations of the extrinsic curvature and the ellipsis represents higher–order or higher–derivative operators. Although these terms do not modify the background evolution at leading order, they can significantly alter the dynamics of scalar perturbations through kinetic mixing (“braiding”) with the metric and through modified gradient structures.

Restoring time diffeomorphism invariance via the Stückelberg field $\pi$, defined by the transformation $t \rightarrow t + \pi(x^\mu)$, and expanding to quadratic order in perturbations, one obtains an effective action for the single propagating scalar degree of freedom after integrating out the non-dynamical lapse and shift. The quadratic action can always be written in the form
\begin{equation}
S^{(2)} = \int dt\, d^3x\, a^3 
\left[
A_{\rm eff}(t)\,\dot{\zeta}^2
- \frac{B_{\rm eff}(t)}{a^2}(\nabla \zeta)^2
\right],
\end{equation}
where $\zeta$ is the gauge–invariant curvature perturbation. The coefficients $A_{\rm eff}$ and $B_{\rm eff}$ depend on the full set of EFT functions, the Hubble rate, and the matter content.

In this general setting, the physical sound speed is defined invariantly as
\begin{equation}
c_s^2 = \frac{B_{\rm eff}(t)}{A_{\rm eff}(t)}.
\end{equation}
The sound speed therefore emerges as the ratio of gradient energy to kinetic energy in the scalar sector. The operator $M_2^4(\delta g^{00})^2$ primarily enhances $A_{\rm eff}$, increasing the kinetic weight of fluctuations and reducing $c_s^2$. Braiding operators proportional to $\delta g^{00}\delta K$ and curvature combinations can modify both $A_{\rm eff}$ and $B_{\rm eff}$ through mixing with metric perturbations.

The simple expression in Eq.~\eqref{basecs_new} is recovered when $(\delta g^{00})^2$ dominates the scalar dynamics and mixing effects are subleading, as occurs in the decoupling limit or in scenarios where braiding operators are independently constrained by gravitational wave and large–scale structure data. In this regime, $M_2^4$ provides a transparent and physically intuitive parameterization of deviations from canonical dark energy.

From an observational standpoint, probes sensitive to the real–time evolution of gravitational potentials on horizon scales are directly sensitive to the effective sound speed and therefore to $M_2^4$. In contrast, background measurements constrain only $\Lambda(t)$ and $c(t)$. Direct measurements of time–dependent scalar metric perturbations are therefore essential to access this otherwise hidden sector of the EFT parameter space. This motivates the use of precision interferometric measurements of horizon–scale strain, which can probe the kinetic operator governing dark energy perturbations and significantly improve constraints on their propagation and clustering properties.

\section{Horizon-Scale Metric Fluctuations as an EFT Probe}

At the perturbative level, dark energy can be described by an equation of state $w$ and a rest-frame sound speed $c_s^2$, which controls the response of pressure perturbations to density fluctuations. Combining energy–momentum conservation with the Einstein equations, the linear evolution of the dark energy density contrast $\delta_{\rm DE}$ in Fourier space obeys schematically
\begin{equation}
\ddot{\delta}_{\rm DE} + (1-3w)H\dot{\delta}_{\rm DE}
+ \left(\frac{c_s^2 k^2}{a^2} - 4\pi G\rho\right)\delta_{\rm DE}
= S(k,t),
\end{equation}
where $S(k,t)$ encodes metric and matter source terms. The term proportional to $c_s^2 k^2/a^2$ represents pressure support, while gravity drives clustering. The competition between these effects defines a Jeans scale,
\begin{equation}
k_J \sim \frac{aH}{c_s},
\end{equation}
which separates regimes of smooth and clustering behavior. For $k \gg k_J$, pressure gradients suppress perturbations and dark energy remains effectively homogeneous. For $k \lesssim k_J$, pressure support is inefficient and dark energy can cluster gravitationally. When $c_s^2 \ll 1$, the Jeans scale approaches the Hubble scale, allowing dark energy perturbations to survive and evolve on horizon-sized modes at late times.

The observational consequences are most transparent in Newtonian gauge,
\begin{equation}
ds^2 = a^2(\tau)\left[-(1+2\Phi)d\tau^2 + (1-2\Psi)d\vec{x}^2\right],
\end{equation}
where $\Phi$ and $\Psi$ denote scalar metric potentials. On large scales, their evolution is governed by the generalized Poisson equation,
\begin{equation}
k^2\Phi + 3a^2H(\dot{\Phi}+H\Phi)
\propto a^2\left(\delta\rho_m + \delta\rho_{\rm DE}\right).
\end{equation}
If $c_s^2 \simeq 1$, dark energy perturbations are suppressed and $\Phi$ decays in the standard manner during cosmic acceleration. In contrast, for $c_s^2 \ll 1$, dark energy clusters on horizon scales, contributing dynamically to the source term and modifying both the amplitude and time dependence of $\Phi$. This constitutes a genuine real-time alteration of metric evolution, rather than a purely line-of-sight integrated effect.

Within the EFT framework, this behavior is controlled primarily by the kinetic operator $M_2^4(\delta g^{00})^2$, which sets the effective sound speed. Small $c_s^2$ (large $M_2^4$) enhances the persistence of horizon-scale potential fluctuations. Since $M_2^4$ does not affect the background expansion at leading order, accessing it observationally requires direct sensitivity to the time evolution of scalar metric perturbations.

A lunar laser interferometer provides precisely such access. Scalar metric fluctuations induce differential redshifts across a macroscopic baseline, generating an effective strain
\begin{equation}
h_{\rm eff}(t) \simeq \frac{\Delta \Phi(t)}{c^2},
\end{equation}
where $\Delta\Phi(t)$ is the potential difference sampled by the interferometer arms. In the ultra low frequency band
\begin{equation}
f \sim 10^{-7} - 10^{-3}\,\mathrm{Hz},
\end{equation}
these measurements correspond to horizon-scale modes, largely inaccessible to terrestrial interferometers and only indirectly probed by large-scale structure or CMB observations. In the frequency domain, the strain power spectrum traces the potential power spectrum at $k \simeq 2\pi f/c$,
\begin{equation}
P_h(f) \propto P_\Phi(k).
\end{equation}
Because $P_\Phi(k)$ at horizon scales is directly influenced by the clustering properties of dark energy, measurements of ultra–low–frequency strain provide sensitivity to the EFT kinetic sector. In this way, a lunar interferometer probes $M_2^4$ not through background expansion or integrated signatures, but through the real-time evolution of scalar gravitational potentials. This makes it orthogonal and complementary to traditional cosmological probes, offering a direct observational handle on the perturbative microphysics of dark energy.

\section{Mock Strain Power Spectra}

To forecast the sensitivity of a lunar laser interferometer to EFT parameters, we construct mock strain power spectra sourced by cosmological scalar perturbations. Throughout, we assume a fiducial $\Lambda$CDM background with $\Omega_m=0.3$, $\Omega_{\rm DE}=0.7$, and $w=-1$, and evolve linear perturbations in the presence of a dark energy component with sound speed $c_s^2$. The evolution of the Newtonian potential $\Phi(k,\tau)$ follows from the linearized Einstein equations and energy–momentum conservation. The qualitative behavior is controlled by the sound horizon (Jeans scale $k_J$). For modes with $k \ll k_J$, pressure gradients are negligible and $\Phi$ remains approximately constant at late times. For $k \gg k_J$, pressure support suppresses dark energy perturbations, leading to oscillatory and decaying solutions for $\Phi$. 

Because the evolution equation is linear, the full solution can be written as
\begin{equation}
\Phi(k,\tau) = \Phi_{\rm prim}(k)\,T_\Phi(k,c_s^2),
\end{equation}
where $\Phi_{\rm prim}(k)$ is the primordial potential generated during inflation and $T_\Phi$ is a transfer function encoding late-time scale dependence. A simple analytic form that reproduces the correct asymptotic limits is
\begin{equation}
T_\Phi(k,c_s^2) \simeq 
\frac{1}{1+\left(\dfrac{c_s k}{aH}\right)^2}.
\end{equation}
For $c_s^2 \sim 1$, clustering is suppressed on horizon scales, while for $c_s^2 \ll 1$ the suppression scale shifts to large $k$, enhancing power at low frequencies.

The potential power spectrum then follows from
\begin{equation}
P_\Phi(k,\tau) = T_\Phi^2(k,c_s^2)\,P_\Phi^{\rm prim}(k),
\end{equation}
where $P_\Phi^{\rm prim}(k)$ is the primordial spectrum. Scalar metric fluctuations induce differential redshifts across an interferometer baseline, producing an effective strain $h \propto \Phi$. In the long-wavelength limit relevant for a lunar detector, the response is coherent across the arms and can be absorbed into an overall normalization. The strain power spectrum is therefore
$P_h(f) = |\mathcal{R}(f)|^2 P_\Phi(k)$, where $\mathcal{R}(f)$ denotes the (orientation-averaged) detector response.

In a recent work \cite{soundspeed}, we derived representative mock spectra for $c_s^2=1$ and $c_s^2=10^{-2}$ over the frequency range $10^{-7}$--$10^{-3}\,\mathrm{Hz}$, wherein it was shown how the spectra can exhibit approximate power-law behavior set by the primordial spectrum and the large-scale transfer function while at high frequencies ($k\gg k_J$), all curves converge because dark energy perturbations are pressure supported and the potential is dominated by matter fluctuations. At low frequencies near the horizon scale, however, small $c_s^2$ enhances clustering, producing significant excess power in $P_h(f)$. 

This scale dependent transition is the key observable signature in our current investigation as well as ultra–low–frequency strain directly traces the sound-horizon imprint in $T_\Phi$, and therefore provides sensitivity to the EFT kinetic parameter $M_2^4$ that controls $c_s^2$. A lunar laser interferometer operates precisely in this regime, making it uniquely suited to constrain dark energy microphysics through horizon-scale metric fluctuations.

\section{Forecast Methodology}

We now describe the statistical framework used to quantify the sensitivity of a lunar laser interferometer to the sound speed of dark energy and the underlying EFT parameters. The primary observable is the strain power spectrum, $P_h(f)$, which is determined by the scalar potential power spectrum and therefore by the EFT kinetic sector. For a parameter vector 
$\boldsymbol{\theta} = \{c_s^2, w, \Omega_{\rm DE}, \ldots\}$,
the theory predicts a strain spectrum
$P_h^{\rm th}(f;\boldsymbol{\theta})$. 
The mapping from interferometric observables to EFT parameters proceeds as follows:
EFT coefficients (such as $M_2^4$) determine the effective sound speed $c_s^2$ as described in previous sections; 
$c_s^2$ fixes the transfer function $T_\Phi(k,c_s^2)$ governing horizon-scale metric fluctuations; 
and $T_\Phi$ sets the frequency dependence of $P_h(f)$ through $k=2\pi f/c$. 
The likelihood therefore directly connects measured strain power to the EFT kinetic operator.

Assuming Gaussian, stationary fluctuations, the likelihood for an estimated spectrum $\hat{P}_h(f)$ is approximated as Gaussian in power-spectrum space,
\begin{equation}
\ln \mathcal{L}(\boldsymbol{\theta})
=
-\frac{1}{2}
\sum_f
\frac{
\left[\hat{P}_h(f)-P_h^{\rm th}(f;\boldsymbol{\theta})\right]^2
}{
\sigma_h^2(f)
},
\end{equation}
where $\sigma_h(f)$ denotes the variance in each frequency bin, determined by instrumental noise and observation time. To forecast parameter sensitivity prior to data acquisition, we use the Fisher information matrix,
\begin{equation}
F_{ij}
=
\sum_f
\frac{1}{\sigma_h^2(f)}
\frac{\partial P_h^{\rm th}(f)}{\partial \theta_i}
\frac{\partial P_h^{\rm th}(f)}{\partial \theta_j},
\end{equation}
evaluated at a fiducial cosmology. The inverse Fisher matrix approximates the parameter covariance,
$\mathrm{Cov}(\theta_i,\theta_j) \simeq (F^{-1})_{ij}$,
so that the marginalized uncertainty is $\sigma(\theta_i)=\sqrt{(F^{-1})_{ii}}$. The Fisher matrix serves three purposes: (i) forecasting the precision with which $c_s^2$ can be measured; (ii) quantifying the statistical significance of deviations from the canonical value $c_s^2=1$; and (iii) identifying regions of EFT parameter space that can be excluded at a given confidence level.

Given the absence of a finalized detector design, we adopt a design-agnostic treatment of uncertainties. In the ultra–low–frequency regime of interest, scalar-induced strain is coherent across the baseline, and we approximate the variance using the standard Gaussian expression with an effective number of independent modes per bin. This signal-focused approach isolates the intrinsic cosmological information content of horizon-scale strain measurements and yields conservative, implementation-independent forecasts. More detailed noise modeling can be incorporated straightforwardly once a specific lunar interferometer configuration is defined, without altering the conceptual mapping between interferometric observables and EFT parameters established here.

\begin{figure}[] \includegraphics[width=0.45\textwidth, keepaspectratio]{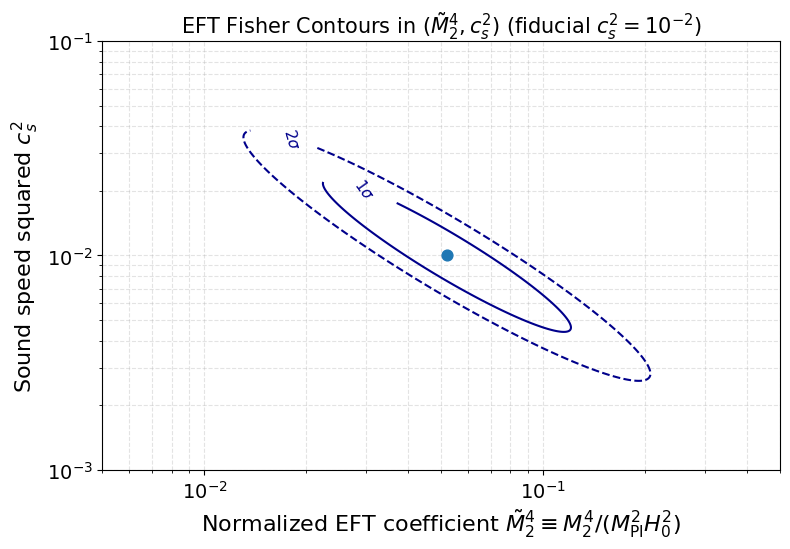}
\caption{Fisher contour ellipses in the $(\tilde{M}_2^{4},c_s^2)$ plane for a lunar laser interferometer, centered on the fiducial clustering dark energy model $(w,c_s^2)=(-1,10^{-2})$. Shown are the joint 1$\sigma$ and 2$\sigma$ confidence regions obtained using the cosmology-calibrated mock strain power spectrum mapped onto the EFT phase space. $\Lambda(t)$ was chosen to reproduce the desired $\Lambda$CDM-like evolution, and the braiding and extrinsic-curvature coefficients were held fixed, consistent with existing gravitational-wave propagation and large-scale structure bounds.}
\label{fig:fisher1e-2_M24_vs_cs2}
\end{figure}

\section{EFT Constraints and Parameter Space}
Figure~1 shows the illustrative Fisher contour ellipses in the $(\tilde{M}_2^{4},c_s^2)$ plane centered on the fiducial model with $c_s^2=10^{-2}$, including the 1$\sigma$ and 2$\sigma$ joint confidence regions. The parameter $\tilde{M}_2^{4}$ is the standard dimensionless EFT normalization $\tilde{M}_2^{4} = \frac{M_2^4}{M_{\rm Pl}^2 H_0^2}$. 
The elongated shape of the Fisher contours in the $(\tilde{M}_2^{4},c_s^2)$ plane reflects a partial degeneracy between the kinetic normalization of the scalar degree of freedom and its effective propagation speed. In the unitary-gauge EFT, the operator $M_2^4(\delta g^{00})^2$ modifies the kinetic coefficient of the Goldstone mode associated with broken time diffeomorphisms, while $c_s^2$ is determined by the ratio of gradient to kinetic terms in the quadratic action for scalar perturbations. When the braiding and extrinsic-curvature operators are fixed to values independently constrained by gravitational-wave and large-scale structure data, varying $\tilde{M}_2^{4}$ primarily rescales the kinetic normalization, whereas $c_s^2$ controls the location of the spectral turnover associated with dark energy clustering. Because the observable strain spectrum is most sensitive to this turnover scale, a change in $\tilde{M}_2^{4}$ can partially compensate a shift in $c_s^2$ by adjusting the overall amplitude of the scalar response without substantially altering the frequency at which the spectrum departs from the smooth-$\Lambda$CDM limit. This compensation produces the characteristic elongated contour, indicating that the experiment constrains a particular linear combination of EFT parameters more tightly than each parameter individually.
\begin{figure}[]   \includegraphics[width=0.45\textwidth, keepaspectratio]{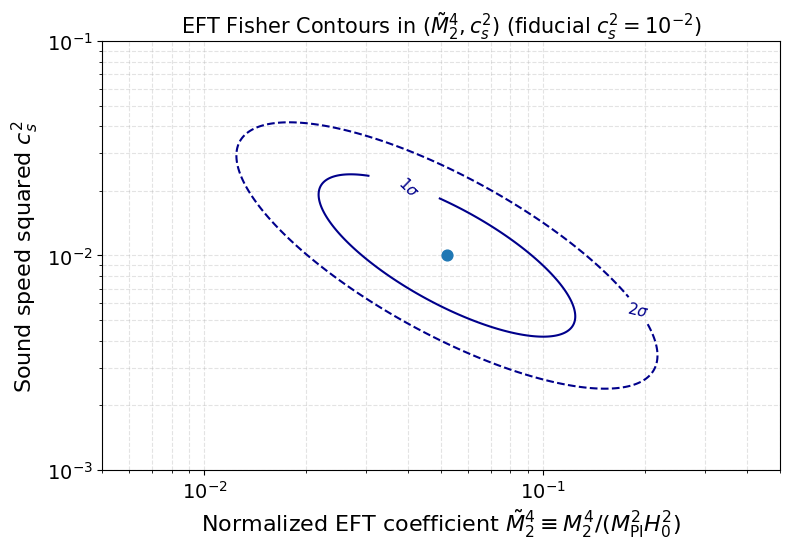}
\caption{Fisher contour ellipses in the $(\tilde{M}_2^{4},c_s^2)$ plane for a lunar laser interferometer, centered on the fiducial clustering dark energy model $(w,c_s^2)=(-1,10^{-2})$. Shown are the joint 1$\sigma$ and 2$\sigma$ confidence regions obtained using the cosmology-calibrated mock strain power spectrum mapped onto the EFT phase space. A Gaussian prior is assumed for the equation-of-state parameter, restricting it to within 3\% of $w=-1$, the braiding parameter is allowed to vary freely, and no additional constraints are imposed on the parameters governing the background expansion history.}
\label{fig:fisher1e-2_M24_vs_cs2_broad}
\end{figure}
The Fisher contours shown in Figure~1 were generated assuming the background expansion was fixed to $w\simeq -1$, while $\Lambda(t)$ was chosen to reproduce the desired $\Lambda$CDM-like evolution. The braiding and extrinsic-curvature coefficients were held fixed, consistent with existing gravitational-wave propagation and large-scale structure bounds. The orientation of the degeneracy direction is directly influenced by braiding. In the general EFT, the scalar sound speed can schematically be written as $c_s^2 = \frac{B_{\rm eff}(t)}{A_{\rm eff}(t)}$
where $B_{\rm eff}$ and $A_{\rm eff}$ denote effective gradient and kinetic coefficients. Nonzero braiding mixes metric and scalar fluctuations, modifying both numerator and denominator and thereby rotating the principal degeneracy direction in parameter space. By fixing braiding parameters, the contours shown here isolate the degeneracy associated purely with the kinetic normalization versus propagation speed. 

The Fisher contour analysis shown in Fig.~2 has been updated to include a Gaussian prior on the dark energy equation-of-state parameter, restricting it to within 3\% of $w=-1$. This prior reflects one of the tightest commonly cited constraints from a single, coherent global analysis in the final \emph{Planck} 2018 cosmological-parameter release, when combined with baryon acoustic oscillations and Type Ia supernovae data~\cite{Planck2018}. Additionally, the braiding parameter is allowed to vary freely, and no constraints are imposed on the parameters governing the background expansion history. Consequently, all EFT parameters are treated as free, which generically broadens the sensitivity contours and increases their tilt, reflecting enhanced degeneracies and compensatory effects between mixing and sound-speed contributions. 

The finite width of the contours along the $c_s^2$ direction directly encodes the interferometer’s sensitivity to the clustering properties of dark energy. The fact that the $1\sigma$ and $2\sigma$ regions occupy only a restricted range in $c_s^2$ demonstrates that the measurement is not merely sensitive to an overall amplitude rescaling, but to the dynamical propagation of the scalar perturbations themselves. In particular, models with $c_s^2 \ll 1$ generate a distinct low-frequency enhancement in the strain spectrum that cannot be fully mimicked by varying $\tilde{M}_2^{4}$ alone. Consequently, a clustering fiducial model (e.g., $c_s^2=10^{-2}$) lies well separated from the canonical smooth dark energy limit $c_s^2=1$ under the assumed signal dominated conditions, indicating that strongly clustering dark energy would be statistically distinguishable from the smooth $\Lambda$CDM case.

Most importantly, the contour geometry demonstrates that the lunar interferometer forecast is genuinely probing the kinetic sector of the EFT rather than simply its background expansion parameters. The constraints on $\tilde{M}_2^{4}$ imply a direct bound on the coefficient of the $(\delta g^{00})^2$ operator, thereby placing limits on the kinetic normalization of the Goldstone mode associated with broken time translations. The comparatively tighter constraint along the $c_s^2$ direction shows that the experiment is especially sensitive to the propagation speed of dark energy fluctuations, establishing that interferometric strain measurements provide a direct observational handle on the microphysical structure of the dark energy EFT. In this sense, the contours represent not merely a phenomenological constraint, but a quantitative bound on the underlying kinetic operators governing scalar perturbations in the effective theory. 
\begin{figure}[] 
  \includegraphics[width=0.45\textwidth, keepaspectratio]{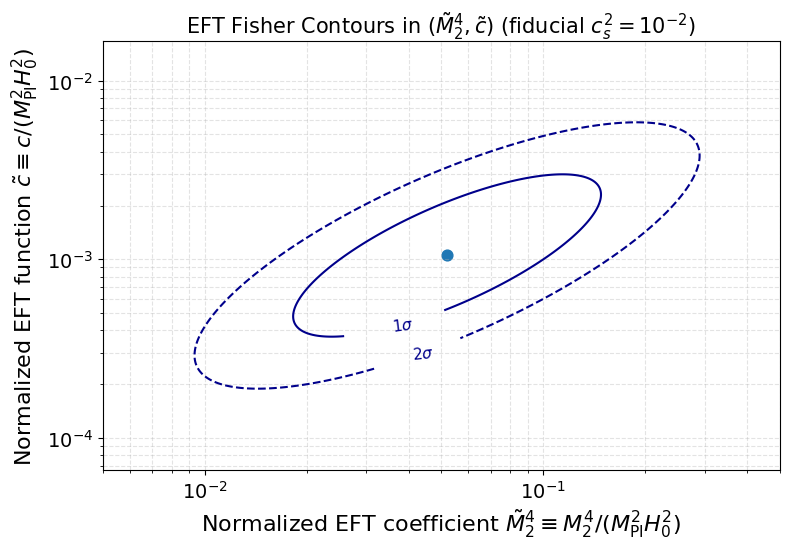}
\caption{Fisher contour ellipses in the $(\tilde{M}_2^{4},c_s^2)$ plane for a lunar laser interferometer, centered on the fiducial clustering dark energy model $(w,c_s^2)=(-1,10^{-2})$, with all EFT parameters allowed to vary freely in the likelihood analysis. Shown are the joint 1$\sigma$ and 2$\sigma$ confidence regions obtained using the cosmology-calibrated mock strain power spectrum mapped onto the EFT phase space.}
\label{fig:fisher1e-2_M24_vs_c(t)}
\end{figure}

Figure~3 presents a corresponding Fisher contour plot in the $(\tilde{M}_2^{4},\tilde{c})$ plane, where $\tilde{c} \equiv \frac{c(t)}{M_{\rm Pl}^2 H_{0}^2}$, centered on the fiducial clustering dark energy model $(w,c_s^2)=(-1,10^{-2})$, with all EFT parameters allowed to vary freely in the likelihood analysis. The elongated shape of the Fisher contours in the $(\tilde{M}_2^{4},\tilde{c})$ plane reflects the intrinsic relation between the background EFT function $\tilde{c}$ and the kinetic operator $\tilde{M}_2^{4}$ that together determine the scalar sound speed. In the unitary-gauge EFT, $\tilde{c}$ sets the normalization of $\rho_{\rm DE}+p_{\rm DE}$ and therefore controls the background departure from exact de~Sitter expansion, while $\tilde{M}_2^{4}$ modifies the kinetic coefficient of scalar perturbations. Because the observable strain spectrum is primarily sensitive to the combination that fixes the clustering scale, variations in $\tilde{c}$ can be partially compensated by correlated shifts in $\tilde{M}_2^{4}$ that preserve $c_s^2$, leading to a positively correlated and elongated contour. 

The finite width of the contours along the $\tilde{c}$ direction demonstrates that the interferometric measurement does not merely constrain the sound speed, but also places a direct bound on the background EFT coefficient that controls the normalization of scalar fluctuations. In particular, the restricted extent of the $1\sigma$ and $2\sigma$ regions implies that large deviations in $\tilde{c}$—and hence large departures from the canonical $\Lambda$CDM kinetic structure—would produce spectrally distinguishable signatures in the low-frequency strain power. The contour geometry therefore indicates that a lunar interferometer can probe both the kinetic normalization and the background coupling encoded in $\tilde{c}$, providing a quantitative constraint on the operator multiplying $g^{00}$ in the dark energy effective field theory. This establishes sensitivity not only to the propagation speed of fluctuations, but also to the underlying background EFT parameter governing the scalar sector.

\section{Discussion and Conclusions}
We have demonstrated that horizon-scale metric fluctuations provide a direct and model-independent observational probe of the kinetic sector of the effective field theory (EFT) of dark energy. By constraining the operator $M_2^4$ and its associated scalar sound speed $c_s^2$, interferometric measurements access a sector of dark energy physics that is effectively invisible to traditional cosmological probes based solely on background expansion or sub-horizon structure growth. In the unitary-gauge EFT description, $M_2^4$ controls the normalization of the kinetic operator $(\delta g^{00})^2$, while the function $c(t)$ governs the departure from exact de~Sitter evolution. The combination of these coefficients determines the propagation and clustering properties of the scalar degree of freedom associated with broken time translations. 

Our results establish that ultra-low frequency strain measurements can constrain these EFT operators directly, without invoking a specific ultraviolet (UV) completion or committing to a particular microphysical dark energy model. This is a crucial conceptual advance: rather than testing individual scalar-field Lagrangians or phenomenological parameterizations, interferometric observations probe the symmetry structure and operator content of the most general low-energy theory consistent with covariance and broken time diffeomorphisms. The measurement constrains the coefficients of the EFT itself, not merely derived phenomenological quantities.

Importantly, the forecasted contours show that strongly clustering dark energy models ($c_s^2 \ll 1$) are spectrally distinguishable from the canonical smooth limit $c_s^2 = 1$. The ability to bound $M_2^4$ simultaneously demonstrates sensitivity to the kinetic normalization of scalar perturbations, establishing that the observable is not reducible to a background equation-of-state constraint. The degeneracy structure in the $(\tilde{M}_2^{4}, c_s^2)$ and $(\tilde{M}_2^{4}, \tilde{c})$ planes further shows that interferometric measurements probe specific operator combinations predicted by the EFT, enabling quantitative tests of the scalar sector’s internal consistency.

More broadly, this work identifies a qualitatively new avenue for testing the microphysical structure of cosmic acceleration. While cosmic microwave background and large-scale structure surveys constrain the background expansion history and gravitational coupling, horizon-scale interferometric observations probe the dynamical response of spacetime itself to dark energy fluctuations. This complementarity elevates interferometric cosmology from a phenomenological probe to a fundamental test of the EFT describing the late-time Universe.

In summary, we have shown that future lunar-scale interferometers can place decisive constraints on the kinetic operators of dark energy, independently of any specific UV-complete construction. By directly bounding $M_2^4$, $c(t)$, and the associated sound speed, such experiments open an observational window onto the symmetry-breaking structure underlying cosmic acceleration, transforming dark energy from a purely background phenomenon into a testable sector of low-energy effective field theory.

It is also important to discuss in passing what would be the theoretical implications from such constraints. The kinetic operator $M_2^4(t)(\delta g^{00})^2$ and its companions such as $m_3^3(t)\,\delta g^{00}\delta K$ and $\mu_2^2(t)\!\left(\delta K^2-\delta K_{ij}\delta K^{ij}\right)$ are crucial because they determine the perturbation sector microphysics of cosmic acceleration while remaining largely invisible to background probes \cite{eft1gubitosi2013effective,eft2linder2016effective,eft3bloomfield2013dark,eft4liang2023dark,eft5frusciante2014effective}. In the minimal truncation, $M_2^4$ directly controls the scalar sound speed via $c_s^2=c(t)/\!\big(c(t)+2M_2^4(t)\big)$, so a measurement favoring \emph{large} $M_2^4$ implies $c_s^2\!\ll\!1$ would mean that dark energy is kinetically "heavy" and clusters on near-horizon scales while its EFT sits in a strongly non-canonical regime whose UV completion must accommodate large higher order operator weights (and potentially a lower strong coupling scale) while still satisfying stability/consistency conditions. Conversely, if the data drive these coefficients small (like $M_2^4\!\approx\!0$ with weak braiding/curvature operators), then $c_s^2\!\approx\!1$ and the EFT reduces toward the canonical, weakly coupled quintessence-like limit, meaning the derivative expansion is efficiently suppressed and additional $\delta g$ operators are genuinely subleading. This would be a strong hint that the UV completion flows to a simple low energy scalar sector with minimal mixing and little room for hidden perturbative structure \cite{uv1melville2020positivity,uv2narain2018ultraviolet}.

From the EFT viewpoint, the theoretical default expectation is that these operator coefficients are not fixed by the background (since $\Lambda(t)$ and $c(t)$ set that), so without symmetry protection they are naturally allowed to be $\mathcal{O}(M_{\rm Pl}^2H^2)$ in dimensionful units, making direct constraints on their magnitude a decisive discriminator of whether late time acceleration is a near-canonical, UV kind sector or a strongly non-canonical EFT requiring nontrivial UV structure.

\noindent \textbf{Acknowledgements:} We gratefully acknowledge support from Vanderbilt University and the U.S. National Science Foundation. This work is supported in part by NSF Award PHY-2411502 and by the Vanderbilt Discovery Doctoral Fellowship.

\bibliography{references}

\end{document}